\documentclass[11pt]{article}
\usepackage{jcappub} 
\usepackage{color}
\usepackage{latexsym}
\usepackage{graphicx}
\usepackage{subfigure}
\usepackage{hyperref}
\usepackage{amssymb}
\usepackage{bm}
\usepackage{natbib}
\usepackage{amsmath}

\newcommand\prd{Phys.~Rev.~D}
\def\apj{{ApJ}}                 
\def\apjl{{ApJ}}                
\def\physrep{{Phys.~Rep.}}               
\def\jcap{J. Cosmology Astropart. Phys.}
		
\def\mnras{MNRAS}             

\title{High redshift signatures in  the 21~cm forest due to cosmic string wakes}

\author[a]{Hiroyuki Tashiro,}
\author[b,c]{Toyokazu Sekiguchi}
\author[d,e,f]{and Joseph Silk}

\affiliation[a]{Physics Department, Arizona State University, Tempe AZ 85287, USA.}
\affiliation[b]{Department of Physics and Astrophysics, Nagoya
University, Nagoya 464-8602, Japan}
\affiliation[c]{University of Helsinki and Helsinki Institute of Physics, 
P.O. Box 64, FI-00014, Helsinki, Finland}

\affiliation[d]{Institut d'Astrophysique, UMR 7095 CNRS,\\
Universit\'{e} Pierre et Marie Curie,98bis Blvd Arago, 75014 Paris, France}
\affiliation[e]{Department of Physics and Astronomy, \\The Johns Hopkins University, Homewood Campus, Baltimore MD 21218, USA}
\affiliation[f]{Beecroft Institute of Particle Astrophysics and Cosmology, Department of Physics, \\University of Oxford, Oxford OX1 3RH, UK}

\emailAdd{hiroyuki.tashiro@asu.edu}

\abstract{
Cosmic strings induce minihalo formation in the early
universe. The resultant minihalos cluster in string wakes and create
a ``21~cm forest'' against the cosmic microwave background~(CMB) spectrum.
Such a 21~cm forest can contribute to angular fluctuations of 
redshifted 21~cm signals integrated along the line of sight.
We calculate the root-mean-square amplitude of the 21~cm fluctuations due
to strings and show that these fluctuations can dominate signals from
minihalos due to  primordial density fluctuations at high
redshift~($z\gtrsim 10$), even if the string tension is below the
current upper bound, $G \mu < 1.5 \times 10^{-7}$.
Our results also predict that the Square Kilometre Array~(SKA) can potentially detect the
21~cm fluctuations due to strings
with $G \mu \approx 7.5 
\times 10^{-8}$ for the single frequency band case and $4.0 \times
10^{-8}$ for the multi-frequency band case. 
}

\begin{document}
\maketitle

\section{Introduction}

Cosmic strings are linear topological defects that could form at phase
transitions in the early universe~\cite{1976JPhA....9.1387K}~(see refs.~\cite{1994csot.book.....V,1995RPPh...58..477H} for
reviews).  Probing the observational
signatures of cosmic strings therefore allows us to constrain  high energy particle
physics and the early universe.

Since cosmic strings have energies which depend on the energy scale of early phase transitions,
they can gravitationally induce various observational phenomena.
In particular, cosmic strings can produce density fluctuations~\cite{1980MNRAS.192..663Z,1981PhRvL..46.1169V} and
serve as seeds for the formation of large-scale structures and galaxies~\cite{1984PhRvL..53.1700S}. 
Recent cosmological observations have shown that the cosmic string contribution to
the density fluctuations and large-scale structure formation is
sub-dominant~\cite{1997PhRvL..79.4736A,2003PhRvD..68b3506P,2005PhRvD..72b3513W}.
Constraints on the gravitational interaction of cosmic strings are
given in terms of $G \mu$ where $G$ is Newton's constant and
$\mu$ is the string tension.
A current strong constraint on $G \mu$ is obtained from CMB anisotropy observations.
Planck data provide the limit, $G\mu <1.5 \times
10^{-7}$~\cite{2013arXiv1303.5085P}.
However, cosmic strings below the current limit still have the potential 
to induce early structure formation and early
reionization~\cite{2004PhRvD..70f3523P,2006PhRvD..74f3516O,2012JCAP...05..026S,2012PhRvD..85l3535T,2013JCAP...04..045D}.
In particular, moving long cosmic strings can develop virialized planar
objects called  'string wakes' after the epoch of radiation-matter equality.

Observations of redshifted 21~cm lines from neutral hydrogen are expected to be good probes for the signatures of structure formation due to cosmic strings.
The intensity of redshifted 21 cm lines depends on the number density
and temperature of neutral hydrogen.  Therefore, the measurement of
redshifted 21~cm lines enables us to access the evolution of structure
formation and the thermal state of  the intergalactic medium through the
epoch of reionization ($8 \lesssim z \lesssim 20$) to the dark ages ($z
\gtrsim 20$)~(for reviews, see
refs.~\cite{2006PhR...433..181F,2012RPPh...75h6901P}).
Currently, there are several ongoing and future projects for measuring highly
 redshifted 21~cm lines;
MWA\footnote{http://www.haystack.mit.edu/ast/arrays/mwa/},
LOFAR\footnote{http://www.lofar.org/},
GMRT\footnote{http://gmrt.ncra.tifr.res.in},
SKA\footnote{http://www.skatelescope.org/} and Omniscope~\cite{2010PhRvD..82j3501T}.
Several papers have investigated redshifted 21 cm signatures from string
wakes produced by long
strings~\cite{2010JCAP...12..028B,2011JCAP...08..014H,2012JCAP...07..032H,2013JCAP...02..045M}
and filaments created by string loops~\cite{2012JCAP...05..014P,2013PhRvD..87l3535T}.
However, since planar and filamentary objects are  gravitationally unstable,
string wakes and filaments can fragment to ``minihalos'' whose virial
temperature is below $10^4~$K~\cite{2012JCAP...05..026S,2013JCAP...04..045D}.
The minihalos due to string wakes can dominate nonlinear structures at
$z \gtrsim 15$, even if $G \mu$ is set to the current upper bound.

Minihalos are virialized nonlinear objects whose masses span the  range from
less than $10^{4} ~M_\odot$ to $10^{8}~M_\odot$.  Therefore
detection of minihalos is an  important channel for  accessing information about
small-scale density fluctuations.  Since the virial temperature of
minihalos is generally below the threshold for atomic hydrogen line
cooling, and in the lower mass range even below the threshold for
molecular hydrogen cooling, they cannot further collapse to form
stars. Accordingly, it is difficult to observe minihalos directly
because they are not accompanied by luminous objects.  However, the
hydrogen density and the temperature in minihalos is high enough to
contribute to deviations of the population of the hyperfine levels in
neutral hydrogen from the background values.  Therefore, minihalos can
in principle be detectable as sources of redshifted 21~cm
lines~\cite{2002ApJ...572L.123I, 2002ApJ...579....1F,2006ApJ...646..681S,2011MNRAS.417.1480M}.

In this paper, we study redshifted 21~cm signals from minihalos due to
cosmic strings. Such minihalos cluster in cosmic string wakes and create
a ``21~cm forest'' that is detectable against the CMB spectrum.  Since wakes are
distributed discretely, the 21~cm forest in wakes contributes to the
angular fluctuations of 21~cm signals.
Generally, clustering (or non-Gaussianity) of minihalo distributions
can enhance the fluctuations of 21~cm signals integrated along the line of sight~\cite{2012MNRAS.426L..21C}.
Clustering depends on the fragmentation of wakes to minihalos. Adopting
a simple analytic model, we calculate the fluctuations of 21~cm
signals due to string wakes.

This paper is organized as follows.
In section~\ref{sec:minihalo-0}, we  briefly review the formation of minihalos
due to cosmic strings.
In section~\ref{sec:signal}, we discuss the 21~cm signal from a single minihalo.
In section~\ref{sec:rms}, we calculate observational predictions of  rms fluctuations of
the redshifted 21~cm lines from minihalos due to cosmic strings.
We also discuss the detectability of the fluctuation signals by SKA. 
We conclude in section~\ref{sec:conclusion}.
Throughout the paper, we are using natural units,~$\hbar = c = 1$.
We also assume a flat $\Lambda$CDM model with cosmological parameters,~$h=0.7$, $h^2 \Omega_b=0.023$ and
$h^2 \Omega_{c} = 0.115$.

\section{Minihalo formation in cosmic string wakes} \label{sec:minihalo-0}

Due to the geometrical effects of a moving long string, 
matter accretes onto a planar wake formed behind this string. 
The wake grows by matter accretion and finally becomes a planar virialized object. However,
since a planar object is unstable by virtue of  its own self-gravity, the wake fragments into minihalos.
In this section, following ref.~\cite{2013JCAP...04..045D}, we briefly review  minihalo
formation in cosmic string wakes.

\subsection{Cosmic string wakes} \label{sec:minihalo-1}

The geometry around a straight cosmic string is conical with the deficit
angle, $\alpha = 8 \pi \mu$~\cite{1981PhRvD..23..852V}. Due to this geometrical effect, a
moving long string with relativistic velocity $v_s$
gives matter a kick velocity in the direction of the
plane swept out by the string,
\begin{equation} 
v_k=4 \pi G \mu v_s \gamma_s,
\end{equation}
where $\gamma _s$ is the Lorentz factor of $v_s$.
Accordingly, the two streams of matter overlap
in the string wake and produce an overdense region.
After radiation-matter equality, 
the wake evolves by gravitational instability and finally collapses to
a virialized planar object~\cite{1984PhRvL..53.1700S}.

The evolution of a string wake has been studied by using the Zel'dovich approximation~\cite{1990PhRvD..41.1764P,1994csot.book.....V}.
We consider a matter particle at an initial comoving distance $x$ on a
wake. At the initial time $t_i$, the matter particle obtains a kick
velocity $v_k$.
The trajectory of this particle can be written as
\begin{equation}
 r(x, t) = a(t)(x+\psi(x, t)),
\end{equation}
where $\psi$ is the comoving displacement~(for simplicity, we assume
that the scale factor is normalized as $a(t_i) = 1$ in this section).
In the Zel'dvich approximation, the evolution of $\psi$ in the matter
dominated epoch is given by
\begin{equation}
 \ddot \psi + \frac{4}{3t} \dot \psi -\frac{2}{3 t^2} \psi =0,
\label{eq:phi}
\end{equation}
with the initial condition, $\psi(x,t_i) =0$ and $\dot \psi (x, t_i) =-
v_k \epsilon(x)$ where 
$\epsilon(x ) =1$ for $x>0$ and $\epsilon(x ) =-1$ for $x<0$.
In eq.~(\ref{eq:phi}), the dot denotes the time-derivative.
The solution of eq.~(\ref{eq:phi}) can be written in the limit of $t \gg
t_i$ as
\begin{equation}
   \psi(t) \approx -\frac{3 v_k t_i}{5} 
  \left(\frac{t}{t_i }   \right)^{2/3} .
\label{eq:evo-w}
\end{equation}

The turn-around comoving surface $x_{\rm ta}$ at time $t$, which is the distance of a
matter particle decoupling from the Hubble expansion, is obtained by
the condition $\dot r(x_{\rm ta}, t) =0$.
This condition is equivalent with $x_{\rm ta} (t)+ 2 \psi(x_{\rm ta},
t)=0$. The turn-around condition
provides the turn-around comoving surface as
\begin{equation}
 x_{\rm ta} (t,t_i)= \pm \frac{6}{5} v_k t_i\left( \frac{t}{t_i} \right)^{2/3} .
\end{equation}
From the turn-around condition, $x_{\rm ta} (t)+ 2 \psi(x_{\rm ta}, t)=0$,
the wake width, which is the physical distance from the wake's center to
the turnaround surface, is
\begin{equation}
 w(t,t_i) = |r(x_{\rm ta}, t)| = \frac{1}{2} a(t) x_{\rm ta}(t,t_i).
\label{eq:width}
\end{equation}

After turn-around, accretion proceeds, and accreted matter is finally virialized. We assume that accreted matter reaches a
virialized state when its radius shrinks by a factor of 2 from that
at turnaround. This assumption for  virialization gives the relation
between the virialization time $t_c$ and the turnaround time $t_{\rm
ta}$; $t_c = (1+1/\sqrt{2})^{3/2}t_{\rm ta}$.

Due to  virialization, 
the accreted matter is thermalized in a wake and heated to $T_w$
provided by
\begin{equation}
 T_w = \frac{\mu_m m_p}{3 k_B} 
  v_{c}^2,
\end{equation}
where $k_B$ is the Boltzmann constant, $m_p$ is the proton mass, $\mu_m$ is the mean molecular
weight,
and $v_{c}$ is the velocity of accreted matter at the time $t_c$.
With the relation between $t_{\rm ta}$ and $t_c$, the thermalized temperature is evaluated as
\begin{equation}
T_w
\approx  0.7
 ~ (v_s \gamma_s)^2 \left(\frac{G\mu}{10^{-7}}\right)^2
 \left( \frac{z_i+1}{z+1}\right) ~[{\rm K}],
\label{eq:temperature}
\end{equation}
where 
$z$ and $z_i$ correspond to the redshifts at $t$ and $t_i$, respectively.

\subsection{Minihalo formation} \label{sec:minihalo}

The size of a cosmic string wake depends on the initial time $t_i$ at
which a string starts to produce the wake. Therefore, we need
to model the cosmic string network. The string network is likely to
evolve toward a scaling solution at which the string distribution is
statistically independent of time if string lengths are scaled to the
Hubble radius.
For  a  representation of the cosmic string network, we adopt an analytical toy model given in refs.~\cite{2011JCAP...08..014H,2013JCAP...04..045D}.

Cosmic strings can produce wakes after the time of matter-radiation
equality, $t_{eq}$.
We divide time from the time $t_{eq}$ to the present time $t_0$ into Hubble time steps.
We label the Hubble time steps as $t_1=t_{eq},~t_2=2 t_{eq},...,t_m=2^{m-1}t_{eq}$. 
In each time step, we assume $N_s$ long strings
per Hubble volume. Each string has a length $\gamma t$ ($\gamma \lesssim
1$) and a velocity $v_s$ in a random direction.
We also assume that the string networks are uncorrelated between different
Hubble time steps.

Now we consider a long string laid down at time $t_m$.
We assume that the string makes a virialized wake with the initial dimension $\gamma t_m \times
v_s \gamma_s t_m \times w_m(z_m)$ where $w_m(z) =w(t,t_m) $ and $z_m$ denotes
the redshift at time $t_m$.
A wake evolves with time. The planar dimension $\gamma t_m \times
v_s \gamma_s t_m$ grows due to the
Hubble expansion, while the growth of the wake width $w_m$ follows eq.~(\ref{eq:width}).
Therefore, the volume of the wake at a redshift $z$,
\begin{equation}
V_{\rm wake} (z, z_m) =(\gamma t_m \times v_s \gamma_s t_m \times w_m(z))\left(\frac{z_m+1}{z+1} \right)^2.
\end{equation}
The mass of the wake at $z$ is given by
\begin{equation}
 M_{\rm wake}(z,z_m) = 4\rho_b(z) V_{\rm wake}(z, z_m) ,
\end{equation}
where $\rho_b(z)$ is the background density at $z$ and,  in a factor of four, a factor of two comes from eq.~(\ref{eq:width})
and the another of two is due to the fact that the virial radius is half
of the wake width.

A planar wake is gravitationally unstable. A wake fragments into filamentary
structures and, finally,  these filaments break into beads.
The length scale of the fastest growing mode is roughly $2\pi w(t)$ in both fragmentations to filaments and to beads~\cite{1987PThPh..78.1051M,2010arXiv1002.4870Q, 2012JCAP...05..026S}.
Therefore, we assume that a string wake fragments to beads with radius $w_m(z)$.
Since the density in a virialized wake is four times larger than the
background one,
the mass of a fragmented bead is given by
\begin{equation}
 M_m(z) = \frac{16\pi}{3}  \rho_b(z) w_m^3(z). 
\label{eq:halo_mass}
\end{equation}

Finally,  the resulting beads are virialized to form quasi-spherical halos.
The virial temperature of a virialized halo can be obtained from 
\begin{equation}
T_{\rm vir} = \frac{\mu_m}{3} \frac{m_{p}}{k_{B}} v_{\rm vir}^2,
\end{equation}
where $v_{\rm vir}$ is the gas velocity of the halo.
Assuming that the virialization is accomplished at the  half-radius
of the wake $w$, $r_{\rm vir}= \omega /2$,
we can obtain $v_{\rm vir}$ by solving equations for the virial 
relation,
\begin{equation}
 2 \frac{1}{2}v_{\rm vir}^2 =\frac{3}{5}\frac{G M}{r_{\rm vir}},
  \label{eq:virial_theorem}
\end{equation}
and
the energy conservation between a pre-virialized bead and a
virialized halo,
\begin{equation}
 \frac{1}{2}v_{\rm b}^2-\frac{3}{5}\frac{G M}{w} =
 \frac{1}{2}v_{\rm vir}^2-\frac{3}{5}\frac{G
  M}{r_{\rm vir}},
  \label{eq:energyconservation}
\end{equation}
where $r_{\rm vir}$ is the radius of
a virialized halo and $v_b$ is the thermal velocity in a bead.
Since the fragmentation does not change the thermal energy
significantly, we assume that the velocity $v_b$ corresponds to the wake
temperature $T_w$ given in eq.~(\ref{eq:temperature}).

\begin{figure}[t]
 \begin{minipage}{0.5\hsize}
  \begin{center}
\includegraphics[width=0.9\textwidth]{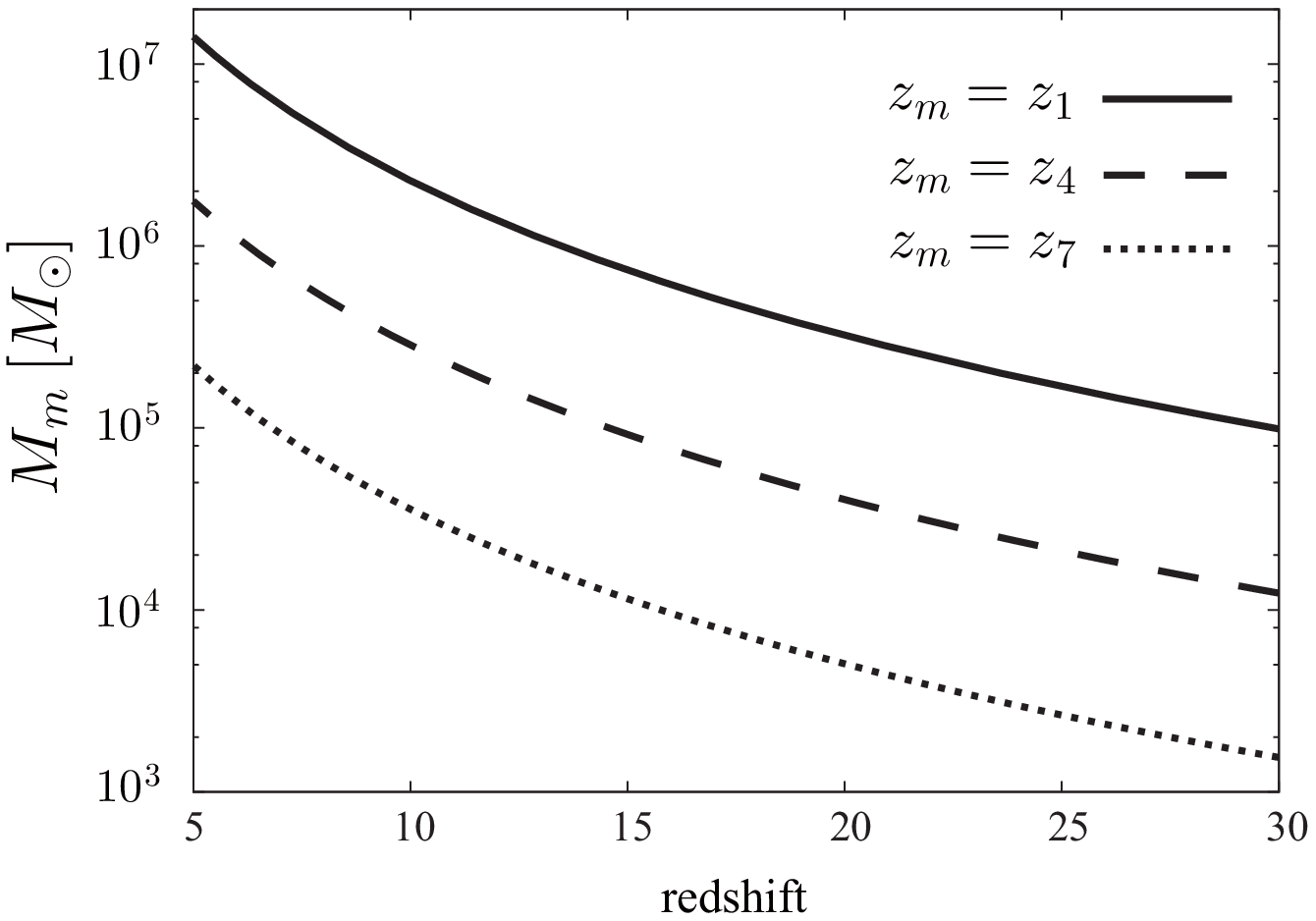}
  \end{center}
 \end{minipage}
 \begin{minipage}{0.5\hsize}
  \begin{center}
\includegraphics[width=0.9\textwidth]{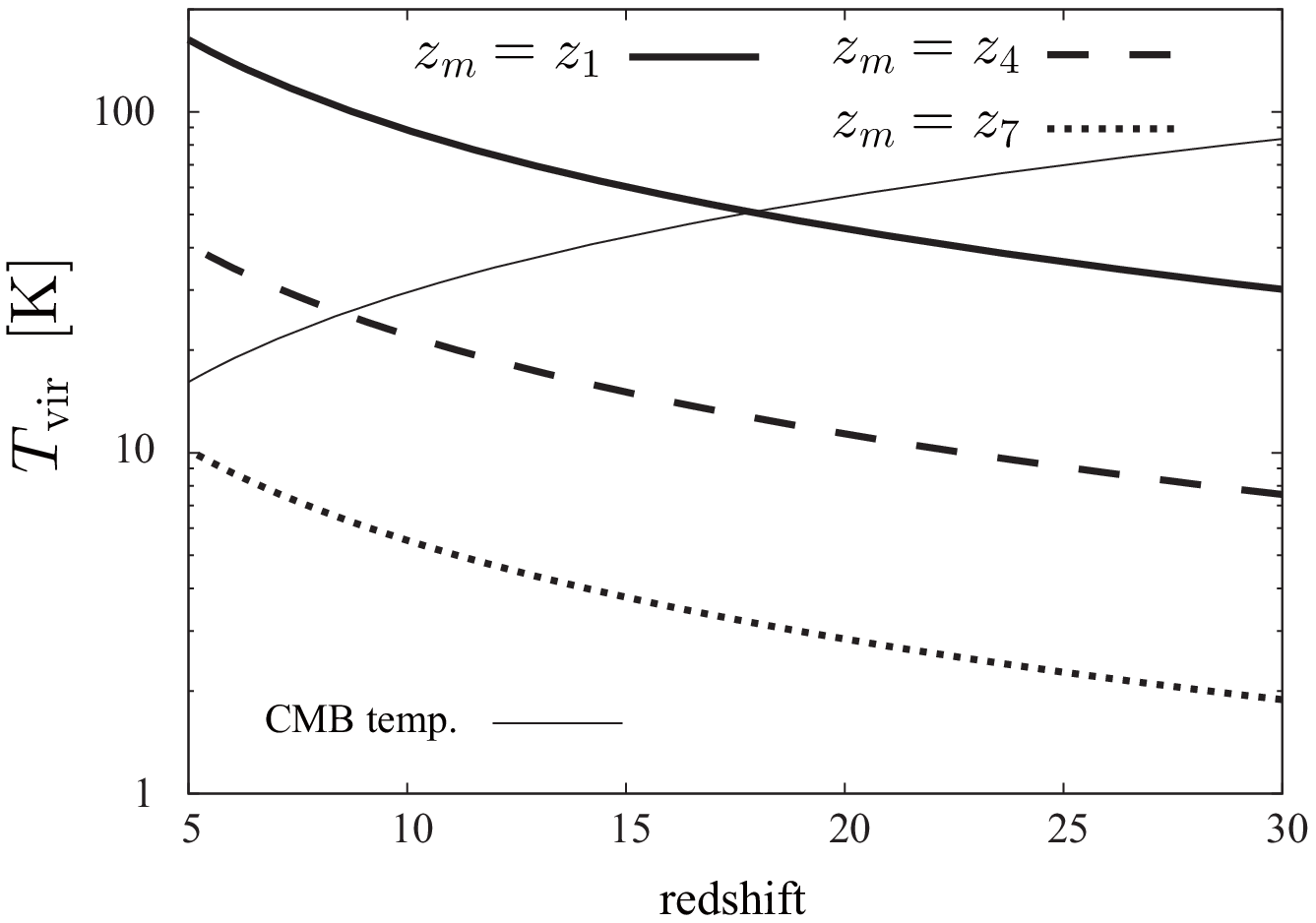}
\end{center}
 \end{minipage}
\caption{Redshift evolutions of halo mass (left panel) and halo virial
 temperature (right panel) in a
 string wake. We set $G \mu = 10^{-7}$.
 In both panels, different types of lines represent 
 different $z_m$ at which a wake formed. The solid, dashed and dotted
 lines are for $z_m = z_1 \sim 3000$, $z_4\sim 1000$ and $z_7 \sim 300$,
 respectively.
In the right panel, the thin line represents the CMB temperature. }
 \label{fig:halo}
\end{figure}

We plot the redshift evolutions of halo mass and virial temperature in 
wakes produced at different $z_m$ in figure \ref{fig:halo}.
In the figure, we
take $G \mu = 10^{-7}$. We also use $v_s = 0.5$ and $\gamma = 1$.
As the redshift decreases, both values increase.
The redshift dependences of halo mass and virial temperature are roughly
$z^{-3}$ and $z^{-1}$.
Halos due to string wakes have lower density and cooler virial temperature
than in the standard $\Lambda$CDM model~(cf.~$T_{\rm vir} \approx 700(M/10^{6} M_\odot)^{2/3} (z/10)$~K~\cite{2001PhR...349..125B}).
As shown in the right panel of figure \ref{fig:halo}, the virial
temperature of halos in wakes is below 200~K even at $z=5$.
The virial temperature of the halos is not high enough 
for atomic cooling~($T_{\rm vir} \gtrsim 10^4$~K) nor even for hydrogen
molecular cooling~($T_{\rm vir} \gtrsim 200$~K).
Therefore, these ``minihalos'' cannot contribute
to star and galaxy formations in the early universe.

For reference, we plot the CMB temperature as the thin line in the right
panel of figure \ref{fig:halo}. At high redshifts, the virial
temperature of halos is lower than the CMB temperature. Therefore, the 21~cm
signals of minihalos are expected as absorption signals against the
CMB. On the other hand, the virial temperature of halos in old wakes at
lower redshifts becomes larger than the CMB temperature.  The signals
from such minihalos are emission signals. We discuss these in more detail in
the following section.

Since we assume that the virial radius is a half of the bead radius, $w(z)/2$,
the typical density of minihalos is eight times larger than the one
of a virialized planar wake. Therefore, in terms of the background matter
density $\rho_b$, the typical density of the minihalos can be
expressed by $\overline{\rho}_{\rm vir} = 32 \rho_b$.

\section{Redshifted 21~cm lines from a minihalo}
\label{sec:signal}

Although a minihalo cannot produce stars and galaxies,
the density of neutral hydrogen becomes denser and the gas
temperature is heated higher than the background values.
As a result,
the spin temperature of a minihalo is decoupled from the CMB temperature
and is different from the background spin temperature.
Therefore, a single minihalo can generate an  observable 21~cm line signal.

The 21~cm signal from a minihalo depends on the profiles of the hydrogen
density and temperature in a minihalo.
The shock thermalization at the virialization of a minihalo makes the minihalo
isothermal at $T_{\rm vir}$. For simplicity, we assume that
the baryon density of a minihalo inside the virial radius is also uniform
due to the shock, $\rho_{\rm vir} =32 \rho_b$.

The 21~cm signal from a single halo is observed as an emission or
absorption signal against
the CMB.
Therefore, it is useful to express the observed 21~cm line flux from a minihalo per unit frequency
relative to the CMB photon flux~\cite{2002ApJ...572L.123I},
\begin{equation}
\frac{d \delta F }{d \nu} =
2{ \nu^2}  {\delta T_b}
  \Delta \Omega_{\rm halo},
  \label{eq:differential_flux_a_halo} 
\end{equation}
where $\delta F$ is the differential total flux from the CMB photon flux,
$\Delta \Omega_{\rm halo}$ is the solid angle subtended by the minihalo,
$\Delta \Omega_{\rm halo} = \pi R_{\rm vir}^2 /D_A^2 $ with the comoving
minihalo radius, $R_{\rm vir} = (1+z) r_{\rm vir}$, and the comoving
angular diameter distance $D_A$ to $z$.
In eq.~(\ref{eq:differential_flux_a_halo}), 
${\delta T_b}$ is the redshifted differential
brightness temperature averaged over the minihalo cross-section, $A = \pi
R_{\rm vir}^2$,   
\begin{equation}
{ \delta T_b } =\frac{1}{1+z}\left(
\frac{\int dA~ T_{b}(\alpha)} {A} -T_{\rm CMB}(z) \right),
\label{eq:ave-tb}
\end{equation}
where $T_{b}(\alpha)$ is the brightness temperature along a line of sight through the
minihalo and a function of impact parameter $\alpha$ (in unit of $R_{\rm vir}$) from the centre of the minihalo.

We obtain $T_{b} (\alpha)$ from the equation,
\begin{equation}
T_b(\alpha)= T_{\rm CMB}(z) e^{-\tau (\alpha)} +\int_0^{\tau (\alpha)} T_S
e^{-\tau'} d\tau' .\label{eq:tbdef}
\end{equation}
where $\tau$ is the 21~cm optical depth of neutral hydrogen to
photons along a line of sight with impact parameter $\alpha$
{and $T_S$ is the spin temperature.}

The 21~cm optical depth with $\alpha$ at frequency $\nu$ is provided by~\cite{2002ApJ...579....1F} 
\begin{equation}
\tau (\nu)= \frac{3 A_{10}T_*}{ 32\pi\nu_*^2  }\int_{-\infty}^{\infty}
 \frac{n_{\rm HI}(\ell)\phi(\nu_{\rm em})}{ T_S(\ell)} dR',
\end{equation}
where $A_{10} =2.85 \times 10^{-15}~ {\rm s}^{-1}$ is the Einstein
$A$-coefficient for the 21 cm transition,
$n_{\rm HI}$ is the number density of neutral hydrogen in the minihalo,
$\phi(\nu)$ is the normalized line profile, $\nu_{\rm em}$ is $\nu_{\rm em} = (1+z) \nu$, 
$\ell$ is a radial comoving distance given by
$\ell^2=R'^2+(\alpha R_{\rm vir})^2$, and the subscript $*$ denotes the
values corresponding to $21~$cm.
Since gas inside the minihalo is thermalized at $T_{\rm vir}$, the line
profile is broadened by the thermal Doppler shift,
{
\begin{equation}
 \phi(\nu_{\rm em})=(\Delta \nu_D \sqrt{\pi})^{-1}\exp \left(
  -\frac{(\nu_{\rm em}-\nu_*)^2}{\Delta \nu_D^2} \right) ,
\end{equation}
}
where $\Delta\nu_D = \nu_* \sqrt{2 k_B T_{\rm vir}/m_{\rm H}}$ with
hydrogen mass $m_{\rm H}$.

The spin temperature {$T_S$} is determined by
the balance between CMB photon absorption, collision between atoms and
scattering off Ly$\alpha$ photons~\cite{1958PIRE...46..240F},
\begin{equation}
 {T_S}=\frac{T_{\rm CMB} + y _c T_{\rm vir} + y_\alpha T_{\alpha}}
  {1+ y_c +y_\alpha},
\end{equation}
where $y_c$ and $y_\alpha$ are the collisional and Ly$\alpha$ coupling
constants. Throughout this paper, we neglect the Ly$\alpha$ coupling terms, because we are
interested in high redshifts where Ly$\alpha$ photon sources~(stars and
galaxies) do not form efficiently. For $y_c$, we use the values in
ref.~\cite{2006ApJ...637L...1K}.

Since the 21~cm signals are either  line emission or absorption,
the total (integrated) differential flux from a minihalo is obtained by
multiplying the flux at $\nu= (1+z) \nu_*$ by a redshifted
effective line width~$\Delta \nu_{\rm eff}$~\cite{2002ApJ...572L.123I},
\begin{equation}
 \delta F  =
{2 \nu^2}  {\delta T_b}
  \Delta \Omega_{\rm halo} {\Delta\nu_{\rm eff}.}
\label{eq:deltaF}
\end{equation}
For the case of a thermal Doppler shift, the effective line width is given by 
$\Delta \nu_{\rm eff} = \sqrt{\pi} {\Delta \nu_D} /(1+z)$.

\begin{figure}[t]
\centering
\includegraphics[width=0.5\textwidth]{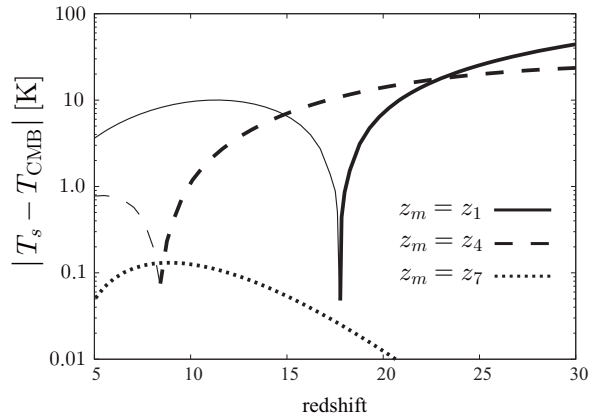}
 \caption{Redshift evolution of the difference between spin temperature
 and CMB temperature in halos inside a string wake.
We set $G\mu = 10^{-7}$.
 The meaning of the 
 line types is the same as in figure~\ref{fig:halo}.
 The thin lines at lower redshifts for $z_m = z_1$ and $z_4$ represent
 positive values (i.e.~emission signals), while the thick lines
 denote negative values (i.e.~absorption signals).}
\label{fig:spin} 
\end{figure}

Since we assume that the gas density and temperature are homogeneous in a
minihalo, the spin temperature of the minihalo is also homogeneous.
According to
eq.~(\ref{eq:tbdef}) in the limit of $\tau \ll 1$,
the differential 21~cm flux from the
minihalo is proportional to ${T_S} - T_{\rm CMB}$. 
We plot the redshift evolution of this value for
wakes with different $z_m$ in figure~\ref{fig:spin}.  
With the gas temperature increasing, the collision coupling becomes
large and the spin temperature is more efficiently dragged to the gas temperature from
the CMB temperature.  The virial temperature of
minihalos is larger in old wakes (small $m$ label) than in young wakes,
as shown in figure~\ref{fig:halo}. Therefore, the deviation of the spin
temperature from the CMB temperature is larger in minihalos inside wakes
with small $m$ label.

Figure~\ref{fig:spin} also shows that the difference between ${T_S}$ and
$T_{\rm CMB}$ becomes large, as
the redshift increases. This is because the virial temperature is low at
high redshifts as shown in figure \ref{fig:halo}.  The collisional coupling at high redshifts is
efficient due to the redshift dependence of the density, even though the virial
temperature decreases.  The spin
temperature is dragged down to the low virial temperature and the
difference from the CMB temperature is enhanced.

As wakes grow, the virial temperature of minihalos increases and, finally, it becomes larger than
the CMB temperature as shown in figure \ref{fig:halo}.
At this time, the 21 cm signal changes from
absorption lines~(${T_S}- T_{\rm CMB} <0$) to emission lines~(${T_S}- T_{\rm CMB} >0$).
We can see this transition for $z_m=z_1$ or $z_4$ in figure~\ref{fig:spin} where the absorption and emission signals are represented
as thick and thin lines, respectively. The redshifts at which the signal
transitions occur correspond to those at which each virial temperature
exceeds the CMB temperature in figure~\ref{fig:halo}.

\section{Fluctuations due to the 21~cm forest in cosmic string wakes}
\label{sec:rms}

Minihalos cluster inside a cosmic string wake. These minihalos create
a ``21~cm forest'' of emission or absorption lines against the CMB
spectrum, when CMB photons pass through a string wakes.
Therefore, the distribution of a 21~cm forest depends on that of the
string wakes and it can contribute to the angular fluctuations
of redshifted 21~cm signals.
In  planning observations by a radio array, the key observable value is the
root-mean-square (rms)
fluctuations of redshifted 21~cm signals integrated along the line of
sight with observation frequency bandwidth.
In this section, we evaluate the rms fluctuations, 
assuming an observation at a mean frequency $\nu_{\rm obs}$ where
the survey volume is pixelized with frequency bandwidth $\Delta \nu$
and beam size $\Delta \theta$ in the longitudinal and transverse directions, respectively

We consider a spherical redshift shell at $z_{\rm obs}$, which satisfies $\nu_{\rm obs}
= (1+z_{\rm obs}) \nu_*$, with width $\Delta z$ corresponding to
the observation frequency bandwidth $\Delta \nu$. 
We divide the shell into cells whose angular size equals  the
beam size $\Delta \theta$ of the observation.
We begin by evaluating the fraction of cells where  a wake is produced by
strings laid down at time $t_m$.

When we consider strings with  number density~$n_s$
and  velocity~$v_s$,
the number of  strings that cross this shell is given by
\cite{2011JCAP...08..014H} 
\begin{equation}
 N_m   =   4\pi r_m^2 n_s  v_s \cos\theta_s/H_m
 =   4\pi r_m^2 N_s  v_s \cos\theta_s H_m^2,
\end{equation}
where $r_m$ is the physical radius of the shell at $z_m$, $H_m$ is the
Hubble parameter at $z_m$ and $\theta_s$ is the angle between the
direction of velocity and the normal vector to the shell. For simply, we
assume the averaged angle is $\langle \cos^2 \theta_s \rangle =1/2$. At the last equality sign, we rewrite $N_m$ in terms of the
number of strings per Hubble volume, $N_s$, for convenience.
The number $N_m$ corresponds to the number of wakes on the shell.

For simplicity, we consider the case where a string vertically
crosses the shell sphere ($\cos \theta_s = 1$). 
The cross section dimension of the string crossing the sphere is
$\gamma t_m \times w_m(z_m)$ at $z_m$. As time
increases, the wake grows and the cross section dimension reaches $\gamma t_m 
(1+z_m)/(1+z_{\rm obs}) \times w_m (z_{\rm obs})$ at $z_{\rm obs}$.
Since the angular size of $w_m(z_{\rm obs})$ much less than $\Delta
\theta$, the wake at $z_{\rm obs }$ stretches
$\gamma t_m  (1+z_m)/(1+z_{\rm obs}) /\Delta d$
cells on the sphere where $\Delta d$ corresponds to the physical scale of $\Delta \theta$ at $z_{\rm obs}$.
Therefore, the fraction of cells at which we can find a wake produced by
strings laid down at time $t_m$ is provided by
\begin{equation}
 f_{{\rm cell },m} = \frac{  (1+z_m) \gamma t_m}
  {  (1+z_{\rm obs}) \Delta d } \frac{N_m}{N_{\rm total}},
\end{equation}
where $N_{\rm total}$ is the total number of cells on the sphere,
$N_{\rm total} =4 \pi /\Delta \Omega_{\rm beam}$ with
$\Delta \Omega_{\rm beam} = \pi (\Delta \theta/2)^2$.
For $f_{{\rm cell},m} \ll 1$, the fluctuations of redshifted $21~$cm lines are caused
by cells occupied by wakes. On the other hand,
for $f_{{\rm cell},m} \sim 1$, the fluctuations are made by void cells (non-occupied cells).
In our parameter region, $f_{{\rm cell},m}$ is below 0.5. Therefore, it
is assumed that the fluctuations are caused by cells occupied by wakes in
this paper.

In a cell occupied by a wake, the volume of the wake at $z_{\rm obs}$ is
$\Delta d \times \Delta l \times w_m(z_{\rm obs})$ where
$\Delta l$ is the
physical scale corresponding to the redshift width $\Delta z$.
As discussed in section~\ref{sec:minihalo},
a wake fragments into beads with radius $w_m(z_{\rm obs})$ and, then,
they are virialized to minihalos with radius $w_m(z_{\rm obs})/2$.
Accordingly, the number of minihalos (beads) in a cell occupied by a wake is
written as
\begin{equation}
 N_{h,m} =\frac{ \Delta d \times \Delta l \times w_m(z_{\rm
  obs})}{4 \pi w_m ^3(z_{\rm obs}) /3}.
\end{equation}
The minihalos are biased tracers of string wakes. Therefore, $N_{h,m}$
corresponds to the linear bias of halos due to the primordial density fluctuations.

So far, we have considered only the case for $\cos \theta_s = 1$.
The different $\theta_s$ modifies the projected cross section on the
observation shell and the occupied volume in a cell. As a result,
$f_{\rm cell}$ and $N_h$ should be functions of $\theta_s$.
However, the averaged direction of strings is $\langle \cos^2 \theta_s \rangle=
1/2$. The difference from the case for $\cos \theta_s =1$
is expected to be a factor of a few. Therefore, we neglect the effect of $\cos \theta_s \neq
1$ on $f_{\rm cell}$ and $N_h$.

We have calculated the differential flux from a single minihalo in
eq.~(\ref{eq:deltaF}). Using this equation, we can write
the differential flux per unit frequency from all minihalos 
in a cell occupied by a wake as
\begin{equation}
 \delta {\cal F}_{\nu,m} = \frac{N_{h,m}}{ \Delta \nu} \delta F_m,
\label{eq:flux-par-frequency}
\end{equation}
where $\delta F_{\nu,m}$ is the differential total flux
of a minihalo induced by a cosmic string laid down at $t_m$.
We define the effective differential brightness
temperature $\overline{\delta T}_{b,m}$ in a cell occupied by a wake as
\begin{equation}
\delta {\cal F}_{\nu,m} \equiv 2\nu^2k_B
\overline{\delta T}_{b,m} \Delta \Omega_{\rm beam}.
\label{eq:defdeltatb}
\end{equation}
According to eqs.~(\ref{eq:deltaF}), (\ref{eq:flux-par-frequency}) and (\ref{eq:defdeltatb}),
we can obtain $\overline{\delta T}_{b,m}$ as
\begin{equation}
\overline{ \delta T} _{b, m} = N_{h,m} \frac{\Delta \Omega_{\rm halo}}{\Delta
 \Omega_{\rm beam}} \frac{\Delta \nu_{\rm eff}}{\Delta \nu} \delta T_{b,m},
\end{equation}
where $\delta T_{b,m}$ is obtained from eq.~(\ref{eq:ave-tb}) for wakes
produced at $t_m$.

Since the fraction of cells occupied by a wake is $f_{{\rm cell},m}$,
the fluctuations of the differential brightness temperature induced
by string wakes laid down at $t_m$ can be expressed as 
\begin{equation}
 \langle \delta T_b^2 \rangle_m  = f_{{\rm cell}, m} ~\overline{ \delta T} _{b, m}^2.
\end{equation}

There is no correlation between string networks at different time steps,
$t_m$. Hence, the total rms fluctuations are obtained by taking the
summation of the contributions from all wakes existing at $z_{\rm obs}$,
\begin{equation}
\langle \delta T_b^2 \rangle ^{1/2} =\left[  \sum_m \langle \delta T_b^2 \rangle_m \right]^{1/2} .
\label{eq:total_rms_fluc}
\end{equation}

\begin{figure}[t]
\centering
 \includegraphics[width=0.5\textwidth]{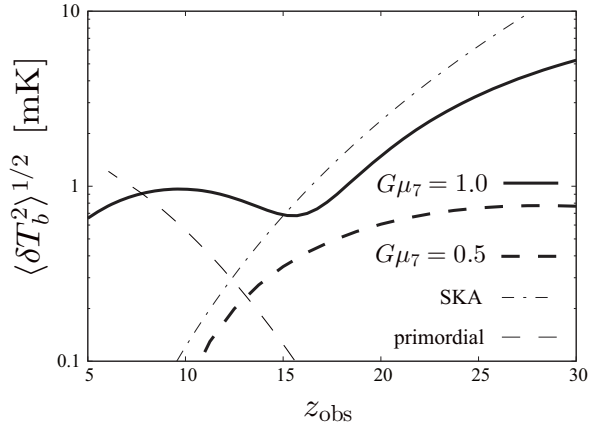}
 \caption{ The rms fluctuations of the differential
 brightness temperature due to string wakes.
The thick solid and dashed lines represent the rms fluctuations for $G
 \mu_7 =1.0$ and $0.5$, respectively, where $G \mu_7$ denotes the
 normalized $G \mu$ in units of $10^{-7}$.
 The SKA noise is plotted as the thin dashed line.}
\label{fig:sqrtTb}
\end{figure}

We show $\langle \delta T_b^2 \rangle ^{1/2} $ for different $G\mu_7
~(G\mu = G\mu_7 \times 10^{-7})$ as
functions of redshift $z_{\rm obs}$ in figure~\ref{fig:sqrtTb}.
Here, we set $\Delta \nu = 1~$MHz and $\Delta \theta =9'$, considering
the current status of SKA.
The
significant contributions to $\langle \delta T_b^2 \rangle ^{1/2} $ are
made by minihalos in wakes produced at earlier times (small $m$ labels).
As shown by the solid lines ($z_m =z_1$) in figure~\ref{fig:spin}, such halos create emission signals
at lower redshifts and absorption signals at higher redshifts. As a
result, most of the 21~cm fluctuations are measured as emission
signals at lower redshifts~($z_{\rm obs} < 15$) for $G\mu_7 =1.0$,
while the absorption signals dominate at higher redshifts~($z_{\rm obs}
> 15$). For $G\mu_7 = 0.5$, the gravitational interaction of cosmic
strings is small so that the strings cannot create halos whose spin temperature is larger
than the CMB temperature at lower redshifts. Therefore, absorption
signals are dominant even at $z_{\rm obs} \gtrsim 10$
and the emission signals are smaller than 0.1~mK even at lower redshifts
for small $G \mu$.

Minihalos are also formed from primordial density fluctuations.
These halos can create the observable fluctuations of redshifted 21~cm lines
for future observations.
In figure \ref{fig:sqrtTb},
we plot the expected $\langle \delta T_b^2 \rangle ^{1/2} $ due to the primordial
minihalos, following ref.~\cite{2002ApJ...572L.123I} with a
Sheth-Tormen mass function~\cite{1999MNRAS.308..119S} and a ``truncated
isothermal sphere'' halo model~\cite{1999MNRAS.307..203S,2001MNRAS.325..468I}.
Since the mass function of primordial minihalos overwhelms that of wake minihalos, which we study in
this paper, at low redshifts, the fluctuations due to the primordial
minihalos are dominant at $z \lesssim 8$.
However, wake minihalos are strongly clustered in a
wake. Accordingly, even though the mass function of string halos is
subdominant around $8 \lesssim z \lesssim 20$~(see figure~2 in ref.~\cite{2013JCAP...04..045D}), 
the fluctuations of redshifted 21~cm lines by wake minihalos become
larger than the ones by primordial minihalos.

In order to illustrate the detectability of the 21~cm signals, we plot the
noise for SKA. The telescope noise for a radio interferometer is given by~\cite{2006PhR...433..181F}
\begin{equation}
 \Delta T_N \approx
0.2~\left(\frac{A_{\rm eff}}{10^6 ~{\rm m}^2}\right)^{-1}
  \left(\frac{10'}{\Delta \theta }\right)^2
    \left(\frac{1+z}{10 }\right)^{4.6}
        \left(\frac{{\rm MHz}}{\Delta \nu } \frac{100 {\rm hr}}{t_{\rm int} }\right)^{1/2}~[{\rm mK}],
\end{equation}
where $A_{\rm eff}$ is the effective collecting area and $t_{\rm int}$ is the integration
time.  In this equation, the system temperature for an observation is
set to the sky temperature of Galactic synchrotron radiation at high
latitudes.  For SKA sensitivity, we adopt $A_{\rm eff} = 10^6~{\rm m}^2$ and $t_{\rm int} =1000~$hr.
Due to the strong frequency dependence of the sky temperature, the SKA
noise grows as $z_{\rm obs}$ increases. However, figure~\ref{fig:sqrtTb} shows that 
21~cm emission fluctuations due to string wakes with $G\mu_7 \sim 1.0$ can dominate the noise at low redshifts, $z_{\rm
obs} < 15$.
We find that SKA can detect the fluctuation signals due to string wakes
up to $G\mu_7 \sim 0.75$.
This detection limit can be improved when the beam size $\Delta \theta $
increases. This is because the fluctuation signal $\langle \delta T_b^2
\rangle^{1/2}$ is proportional to $\Delta \theta^{-1/2}$
while the noise $\Delta T_N$ depends on $\Delta \theta^{-2}$.
For the beam size $\Delta \theta = 15$, we find that the fluctuation signals with $G
\mu_7 =0.5 $ can dominate the noise around $z \sim
13$. However, taking too large $\Delta \theta$ leads to $f_{{\rm cell},m} \sim
1$ for which the fluctuation signals start to decrease as mentioned above.

\section{Conclusion}
\label{sec:conclusion}

In this paper, we have evaluated the redshifted 21~cm fluctuations due to
minihalos in cosmic wakes.
Cosmic strings produce string wakes, and wakes fragment to minihalos in
the early universe.
Although minihalos due to strings can dominate nonlinear structures at high redshifts,
the virial temperature is below 200~K in our models.
Accordingly, they cannot contribute to star and galaxy
formation with atomic cooling and even hydrogen molecular cooling.
However, we have shown that they can create 21~cm line signals,
because the density and temperature of neutral hydrogen in minihalos are  high  enough to 
make the spin temperature of minihalos decouple from the CMB temperature
and differ from the background spin temperature.
The signals are emission lines against CMB at low redshifts, while they are
observed as absorption lines.

Since minihalos due to cosmic strings are clustered in string wakes,
they can strongly enhance the angular fluctuations of redshifted 21~cm
signals integrated over the line of sight.  We have evaluated the
root-mean-square fluctuations due to minihalos clustered in string wakes with a toy
model of string networks.  The
amplitude of the fluctuations due to wake minihalos becomes large  as the
redshift increases.  Therefore, these fluctuations can
dominate those due to minihalos induced by  primordial density fluctuations at high
redshifts, even if the cosmic string tension is smaller than the current
upper limit, $G\mu\sim10^{-7}$.  In order to estimate the feasibility of
the constraint on $G\mu$ by SKA, we have compared the 21~cm fluctuations from
minihalos due to string wakes with the SKA noise.  Although the SKA
noise increases rapidly at high redshifts, we have found that the
fluctuations from such minihalos with $G\mu \sim 7.5 \times 10^{-8}$ at $10<z<15$ is potentially
detectable by SKA.

Here we have considered only one frequency band to evaluate
the detection limit on $G\mu$ by SKA. However, SKA will observe a large
range of frequency and many frequency bands will be available. 
A multi-frequency analysis of 21 cm signals can increase the signal to
noise (SN) ratio. We have performed the SN ratio analysis for SKA multi-frequency observation whose redshift range is $7<z<15$.
We have found that, for the multi-frequency observation,
the SN ratio with $G\mu =4.0 \times 10^{-8}$
becomes larger than one.

The rms fluctuations are enhanced by the number density of
minihalos in string wakes. In this paper, we have calculated the rms
fluctuations, making simple assumptions about  the fragmentation of
string wakes.  The number density of minihalos strongly depends on the
process of fragmentation from wakes. For a detailed study of the
fragmentations, numerical simulations are essential.  We will address
these issues in a future paper.

The amplitude of the rms fluctuations due to strings also depends on the
details of the string network.  However, our calculation is based on a
simple toy model of a scaling solution.  Although we have not
considered the case of a network, string loops are also produced in the evolution of string
networks.  The loops can induce early structure formation and
contribute to  the 21~cm fluctuations. To take into account loop
contributions, our future calculations will be applied with a self-consistent
string network model including the loop distributions.

\acknowledgments
We thank Daisuke Yamauchi for useful comments.
HT is supported by the DOE at ASU. 
TS would like to thank Japan Society for the promotion of Science for financial support.
The research of JS has been supported at IAP by  the ERC project  267117 (DARK) hosted by Universit\'e Pierre et Marie Curie - Paris 6   and at JHU by NSF grant
OIA-1124403.

\providecommand{\href}[2]{#2}\begingroup\raggedright


\begin{thebibliography}{10}

\bibitem{1976JPhA....9.1387K}
T.~W.~B. {Kibble}, {\it {Topology of cosmic domains and strings}},  {\em
  Journal of Physics A Mathematical General} {\bf 9} (Aug., 1976) 1387--1398.

\bibitem{1994csot.book.....V}
A.~{Vilenkin} and E.~P.~S. {Shellard}, {\em {Cosmic strings and other
  topological defects}}.
\newblock 1994.

\bibitem{1995RPPh...58..477H}
M.~B. {Hindmarsh} and T.~W.~B. {Kibble}, {\it {Cosmic strings}},  {\em Reports
  on Progress in Physics} {\bf 58} (May, 1995) 477--562

\bibitem{1980MNRAS.192..663Z}
I.~B. {Zeldovich}, {\it {Cosmological fluctuations produced near a
  singularity}},  {\em \mnras} {\bf 192} (Sept., 1980) 663--667.

\bibitem{1981PhRvL..46.1169V}
A.~{Vilenkin}, {\it {Cosmological density fluctuations produced by vacuum
  strings}},  {\em Physical Review Letters} {\bf 46} (Apr., 1981) 1169--1172.

\bibitem{1984PhRvL..53.1700S}
J.~{Silk} and A.~{Vilenkin}, {\it {Cosmic strings and galaxy formation}},  {\em
  Physical Review Letters} {\bf 53} (Oct., 1984) 1700--1703.

\bibitem{1997PhRvL..79.4736A}
A.~{Albrecht}, R.~A. {Battye}, and J.~{Robinson}, {\it {The Case against
  Scaling Defect Models of Cosmic Structure Formation}},  {\em Physical Review
  Letters} {\bf 79} (Dec., 1997) 4736--4739.

\bibitem{2003PhRvD..68b3506P}
L.~{Pogosian}, S.-H.~H. {Tye}, I.~{Wasserman}, and M.~{Wyman}, {\it
  {Observational constraints on cosmic string production during brane
  inflation}},  {\em \prd} {\bf 68} (July, 2003) 023506.

\bibitem{2005PhRvD..72b3513W}
M.~{Wyman}, L.~{Pogosian}, and I.~{Wasserman}, {\it {Bounds on cosmic strings
  from WMAP and SDSS}},  {\em \prd} {\bf 72} (July, 2005) 023513.

\bibitem{2013arXiv1303.5085P}
{Planck Collaboration}, P.~A.~R. {Ade}, N.~{Aghanim}, C.~{Armitage-Caplan},
  M.~{Arnaud}, M.~{Ashdown}, F.~{Atrio-Barandela}, J.~{Aumont},
  C.~{Baccigalupi}, A.~J. {Banday}, and et~al., {\it {Planck 2013 results. XXV.
  Searches for cosmic strings and other topological defects}},  {\em ArXiv
  e-prints} (Mar., 2013) [\href{http://xxx.lanl.gov/abs/1303.5085}{{\tt
  arXiv:1303.5085}}].

\bibitem{2004PhRvD..70f3523P}
L.~{Pogosian} and A.~{Vilenkin}, {\it {Early reionization by cosmic strings
  reexamined}},  {\em \prd} {\bf 70} (Sept., 2004) 063523.

\bibitem{2006PhRvD..74f3516O}
K.~D. {Olum} and A.~{Vilenkin}, {\it {Reionization from cosmic string loops}},
  {\em \prd} {\bf 74} (Sept., 2006) 063516.

\bibitem{2012JCAP...05..026S}
B.~{Shlaer}, A.~{Vilenkin}, and A.~{Loeb}, {\it {Early structure formation from
  cosmic string loops}},  {\em \jcap} {\bf 5} (May, 2012) 26,
  [\href{http://xxx.lanl.gov/abs/1202.1346}{{\tt arXiv:1202.1346}}].

\bibitem{2012PhRvD..85l3535T}
H.~{Tashiro}, E.~{Sabancilar}, and T.~{Vachaspati}, {\it {Constraints on
  superconducting cosmic strings from early reionization}},  {\em \prd} {\bf
  85} (June, 2012) 123535, [\href{http://xxx.lanl.gov/abs/1204.3643}{{\tt
  arXiv:1204.3643}}].

\bibitem{2013JCAP...04..045D}
F.~{Duplessis} and R.~{Brandenberger}, {\it {Note on structure formation from
  cosmic string wakes}},  {\em \jcap} {\bf 4} (Apr., 2013) 45,
  [\href{http://xxx.lanl.gov/abs/1302.3467}{{\tt arXiv:1302.3467}}].

\bibitem{2006PhR...433..181F}
S.~R. {Furlanetto}, S.~P. {Oh}, and F.~H. {Briggs}, {\it {Cosmology at low
  frequencies: The 21 cm transition and the high-redshift Universe}},  {\em
  \physrep} {\bf 433} (Oct., 2006) 181--301.

\bibitem{2012RPPh...75h6901P}
J.~R. {Pritchard} and A.~{Loeb}, {\it {21 cm cosmology in the 21st century}},
  {\em Reports on Progress in Physics} {\bf 75} (Aug., 2012) 086901,
  [\href{http://xxx.lanl.gov/abs/1109.6012}{{\tt arXiv:1109.6012}}].

\bibitem{2010PhRvD..82j3501T}
M.~{Tegmark} and M.~{Zaldarriaga}, {\it {Omniscopes: Large area telescope
  arrays with only NlogN computational cost}},  {\em \prd} {\bf 82} (Nov.,
  2010) 103501, [\href{http://xxx.lanl.gov/abs/0909.0001}{{\tt
  arXiv:0909.0001}}].

\bibitem{2010JCAP...12..028B}
R.~H. {Brandenberger}, R.~J. {Danos}, O.~F. {Hern{\'a}ndez}, and G.~P.
  {Holder}, {\it {The 21 cm signature of cosmic string wakes}},  {\em \jcap}
  {\bf 12} (Dec., 2010) 28, [\href{http://xxx.lanl.gov/abs/1006.2514}{{\tt
  arXiv:1006.2514}}].

\bibitem{2011JCAP...08..014H}
O.~F. {Hern{\'a}ndez}, Y.~{Wang}, R.~{Brandenberger}, and J.~{Fong}, {\it
  {Angular 21 cm power spectrum of a scaling distribution of cosmic string
  wakes}},  {\em \jcap} {\bf 8} (Aug., 2011) 14,
  [\href{http://xxx.lanl.gov/abs/1104.3337}{{\tt arXiv:1104.3337}}].

\bibitem{2012JCAP...07..032H}
O.~F. {Hern{\'a}ndez} and R.~H. {Brandenberger}, {\it {The 21 cm signature of
  shock heated and diffuse cosmic string wakes}},  {\em \jcap} {\bf 7} (July,
  2012) 32, [\href{http://xxx.lanl.gov/abs/1203.2307}{{\tt arXiv:1203.2307}}].

\bibitem{2013JCAP...02..045M}
E.~{McDonough} and R.~H. {Brandenberger}, {\it {Searching for signatures of
  cosmic string wakes in 21cm redshift surveys using Minkowski Functionals}},
  {\em \jcap} {\bf 2} (Feb., 2013) 45,
  [\href{http://xxx.lanl.gov/abs/1109.2627}{{\tt arXiv:1109.2627}}].

\bibitem{2012JCAP...05..014P}
M.~{Pagano} and R.~{Brandenberger}, {\it {The 21 cm signature of a cosmic
  string loop}},  {\em \jcap} {\bf 5} (May, 2012) 14,
  [\href{http://xxx.lanl.gov/abs/1201.5695}{{\tt arXiv:1201.5695}}].

\bibitem{2013PhRvD..87l3535T}
H.~{Tashiro}, {\it {21 cm angular spectrum of cosmic string loops}},  {\em
  \prd} {\bf 87} (June, 2013) 123535,
  [\href{http://xxx.lanl.gov/abs/1305.4779}{{\tt arXiv:1305.4779}}].

\bibitem{2002ApJ...572L.123I}
I.~T. {Iliev}, P.~R. {Shapiro}, A.~{Ferrara}, and H.~{Martel}, {\it {On the
  Direct Detectability of the Cosmic Dark Ages: 21 Centimeter Emission from
  Minihalos}},  {\em \apjl} {\bf 572} (June, 2002) L123--L126.

\bibitem{2002ApJ...579....1F}
S.~R. {Furlanetto} and A.~{Loeb}, {\it {The 21 Centimeter Forest: Radio
  Absorption Spectra as Probes of Minihalos before Reionization}},  {\em \apj}
  {\bf 579} (Nov., 2002) 1--9.

\bibitem{2006ApJ...646..681S}
P.~R. {Shapiro}, K.~{Ahn}, M.~A. {Alvarez}, I.~T. {Iliev}, H.~{Martel}, and
  D.~{Ryu}, {\it {The 21 cm Background from the Cosmic Dark Ages: Minihalos and
  the Intergalactic Medium before Reionization}},  {\em \apj} {\bf 646} (Aug.,
  2006) 681--690.

\bibitem{2011MNRAS.417.1480M}
A.~{Meiksin}, {\it {The micro-structure of the intergalactic medium - I. The 21
  cm signature from dynamical minihaloes}},  {\em \mnras} {\bf 417} (Oct.,
  2011) 1480--1509, [\href{http://xxx.lanl.gov/abs/1102.1362}{{\tt
  arXiv:1102.1362}}].

\bibitem{2012MNRAS.426L..21C}
S.~{Chongchitnan} and J.~{Silk}, {\it {The 21-cm radiation from minihaloes as a
  probe of small primordial non-Gaussianity}},  {\em \mnras} {\bf 426} (Oct.,
  2012) L21--L25, [\href{http://xxx.lanl.gov/abs/1205.6799}{{\tt
  arXiv:1205.6799}}].

\bibitem{1981PhRvD..23..852V}
A.~{Vilenkin}, {\it {Gravitational field of vacuum domain walls and strings}},
  {\em \prd} {\bf 23} (Feb., 1981) 852--857.

\bibitem{1990PhRvD..41.1764P}
L.~{Perivolaropoulos}, R.~H. {Brandenberger}, and A.~{Stebbins}, {\it
  {Dissipationless clustering of neutrinos in cosmic-string-induced wakes}},
  {\em \prd} {\bf 41} (Mar., 1990) 1764--1774.

\bibitem{1987PThPh..78.1051M}
S.~M. {Miyama}, S.~{Narita}, and C.~{Hayashi}, {\it {Fragmentation of
  Isothermal Sheet-Like Clouds. I ---Solutions of Linear and Second-Order
  Perturbation Equations---}},  {\em Progress of Theoretical Physics} {\bf 78}
  (Nov., 1987) 1051--1064.

\bibitem{2010arXiv1002.4870Q}
A.~C. {Quillen} and J.~{Comparetta}, {\it {Jeans Instability of Palomar 5's
  Tidal Tail}},  {\em ArXiv e-prints} (Feb., 2010)
  [\href{http://xxx.lanl.gov/abs/1002.4870}{{\tt arXiv:1002.4870}}].

\bibitem{2001PhR...349..125B}
R.~{Barkana} and A.~{Loeb}, {\it {In the beginning: the first sources of light
  and the reionization of the universe}},  {\em \physrep} {\bf 349} (July,
  2001) 125--238.

\bibitem{1958PIRE...46..240F}
G.~B. {Field}, {\it {Excitation of the Hydrogen 21-CM Line}},  {\em Proceedings
  of the IRE} {\bf 46} (Jan., 1958) 240--250.

\bibitem{2006ApJ...637L...1K}
M.~{Kuhlen}, P.~{Madau}, and R.~{Montgomery}, {\it {The Spin Temperature and 21
  cm Brightness of the Intergalactic Medium in the Pre-Reionization era}},
  {\em \apjl} {\bf 637} (Jan., 2006) L1--L4.

\bibitem{1999MNRAS.308..119S}
R.~K. {Sheth} and G.~{Tormen}, {\it {Large-scale bias and the peak background
  split}},  {\em \mnras} {\bf 308} (Sept., 1999) 119--126.

\bibitem{1999MNRAS.307..203S}
P.~R. {Shapiro}, I.~T. {Iliev}, and A.~C. {Raga}, {\it {A model for the
  post-collapse equilibrium of cosmological structure: truncated isothermal
  spheres from top-hat density perturbations}},  {\em \mnras} {\bf 307} (July,
  1999) 203--224.

\bibitem{2001MNRAS.325..468I}
I.~T. {Iliev} and P.~R. {Shapiro}, {\it {The post-collapse equilibrium
  structure of cosmological haloes in a low-density universe}},  {\em \mnras}
  {\bf 325} (Aug., 2001) 468--482.

\end{thebibliography}

\end{document}